\documentclass{IEEEtran}
\usepackage{cite}
\usepackage{amsmath,amssymb,amsfonts}
\usepackage{graphicx}
\usepackage{textcomp,nicefrac}
\usepackage{amssymb}
\usepackage{subfigure}
\usepackage{siunitx}
\usepackage{xcolor}
\usepackage{multicol}
\usepackage{multirow}
\usepackage{url}
\usepackage[colorlinks,
            linkcolor=blue,       
            anchorcolor=blue,  
            citecolor=blue]{hyperref}

\def\BibTeX{{\rm B\kern-.05em{\sc i\kern-.025em b}\kern-.08em
T\kern-.1667em\lower.7ex\hbox{E}\kern-.125emX}}
\begin{document}
\title{Effects of shallow carbon and deep N++ layer on the radiation hardness of IHEP-IME LGAD sensors}

\author{Mengzhao Li*, Yunyun Fan*, Xuewei Jia, Han Cui, Zhijun Liang, Mei Zhao, Tao Yang, Kewei Wu, Shuqi Li, Chengjun Yu, Bo Liu, Wei Wang, Xuan Yang, Yuhang Tan, Xin Shi, J. G. da Costa, Yuekun Heng, \IEEEmembership{Member, IEEE}, Gaobo Xu, Qionghua Zhai, Gangping Yan, Mingzheng Ding, Jun Luo, Huaxiang Yin, Junfeng Li, Alissa Howard, Gregor Kramberger

\thanks{This work was supported by the National Natural Science Foundation of China (No.11961141014), the State Key Laboratory of Particle Detection and Electronics (No.SKLPDE-ZZ-202001), the Hundred Talent Program of the Chinese Academy of Sciences (Y6291150K2), the CAS Center for Excellence in Particle Physics (CCEPP), the Scientific Instrument Developing Project of the Chinese Academy of Sciences (No.ZDKYYQ20200007).}
\thanks{Mengzhao Li*, Yunyun Fan*, Xuewei Jia, Han Cui, Zhijun Liang, Mei Zhao, Tao Yang, Kewei Wu, Shuqi Li, Chengjun Yu, Bo Liu, Wei Wang, Xuan Yang, Yuhang Tan, Xin Shi, J. G. da Costa, Yuekun Heng are with the Institute of High Energy Physics, Chinese Academy of Sciences, Beijing 10049, China (e-mail: Zhijun Liang (liangzj@ihep.ac.cn), Mei Zhao (zhaomei@ihep.ac.cn)), and Mengzhao Li, Xuewei Jia, Han Cui, Tao Yang, Kewei Wu, Shuqi Li, Chengjun Yu, Yuhang Tan, Yuekun Heng are also with the School of Physical Sciences, University of Chinese Academy of Sciences, Beijing 100049, China.}
\thanks{Gaobo Xu, Qionghua Zhai, Gangping Yan, Mingzheng Ding, Jun Luo, Huaxiang Yin, Junfeng Li are with the Institute of Microelectronics, Chinese Academy of Sciences, Beijing 100029, China.}
\thanks{Alissa Howard, Gregor Kramberger are with the Jozef Stefan Institute, SI-1000 Ljubljana, Slovenia.}}

\maketitle

\begin{abstract}
Low Gain Avalanche Diode (LGAD) is applied for the High-Granularity Timing Detector (HGTD), and it will be used to upgrade the ATLAS experiment.
The first batch IHEP-IME LGAD sensors were designed by the Institute of High Energy Physics (IHEP) and fabricated by the Institute of Microelectronics (IME). 
Three IHEP-IME sensors (W1, W7 and W8) were irradiated by the neutrons up to the fluence of \SI{2.5}{\times10^{15}~n_{eq}/\centi\metre^2} to study the effect of the shallow carbon and deep N++ layer on the irradiation hardness. Taking W7 as a reference, W1 has an extra shallow carbon applied, and W8 has a deeper N++ layer. 
% After neutron irradiation up to \SI{2.5}{\times10^{15}~n_{eq}/\centi\metre^2}, 
% the leakage current of the three sensors has increased from \SI{}{\nano\ampere} level to \SI{}{\micro\ampere} level, 
% the timing resolution of the three sensors can reach 50 ps, 
% and the maximum collected charge of the three sensors are 10.9~fC (W1), 9.3~fC (W7) and 9.5~fC (W8). 
%Compared with W7 and W8, carbon-implanted W1 has a higher collected charge.
The leakage current, the collected charge and timing resolution of the three IHEP-IME sensors mearured from the Bete telescope test all meet the HGTD requirements(\textless 125 \SI{}{\micro\ampere/\centi\metre^2},\textgreater 4 fC and \textless 70~ps after \SI{2.5}{\times10^{15}~n_{eq}/\centi\metre^2} irradiation fluence).
The W1 sensor with shallow carbon is the most radiation hardness, while the W8 sensor with the deep N++ layer showed the worst radiation hardness.
%Compared with W7 and W8, carbon-implanted W1 has a higher collected charge. % and a wider optimal voltage range. 
%Therefore, the carbon implantation significantly improves the radiation hardness of IHEP-IME LGAD sensors.

\end{abstract}


%{highlights}




%%%%
%%%% Introduction
%%%%
\section{Introduction}
\label{sec_intuduction}

\begin{figure}[tbp]
 \begin{center}
\rotatebox{0}{\includegraphics [scale=0.3]{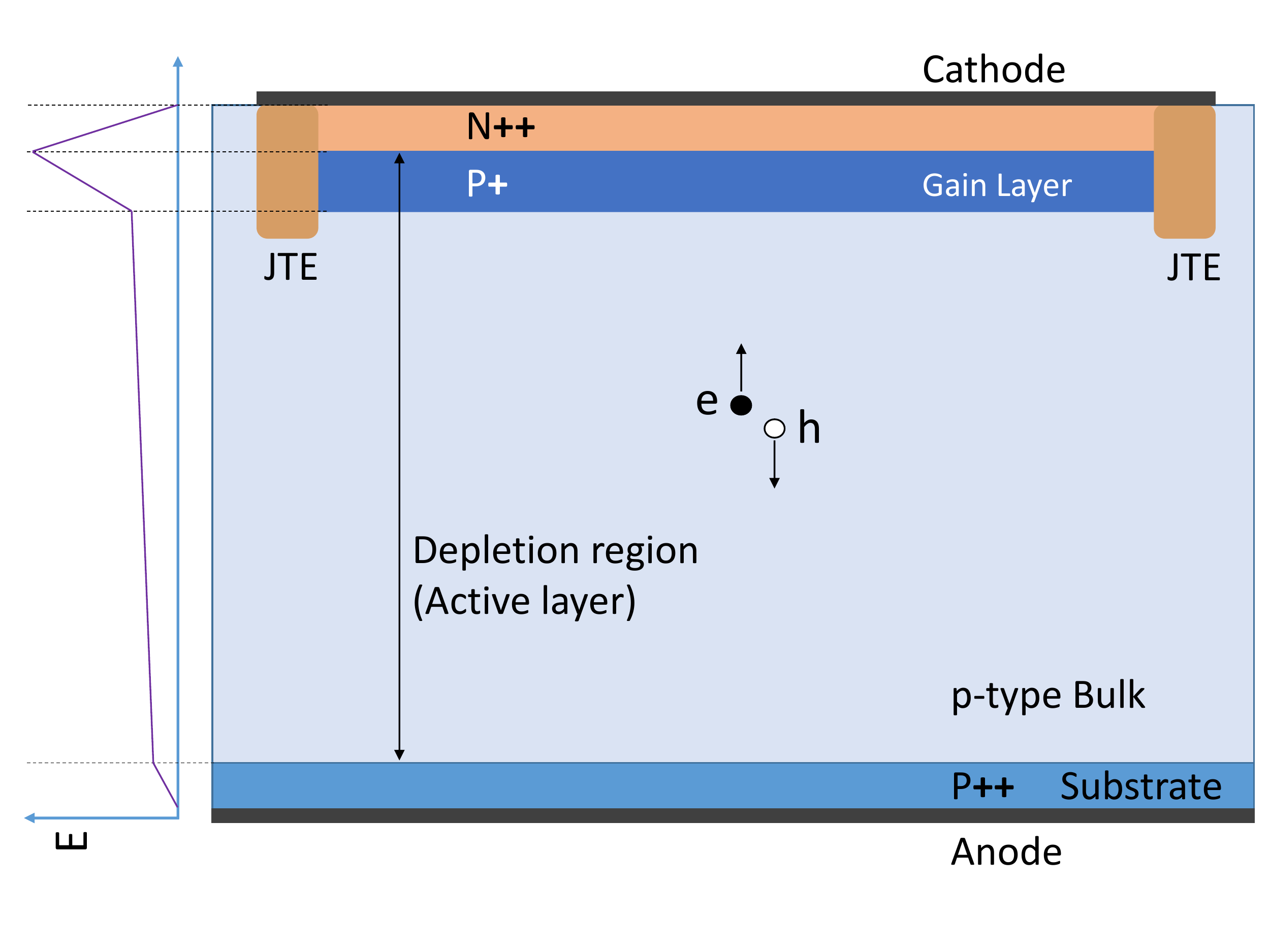}}% insert figure
\caption{Shematic for the LGAD sensors.}
\label{fig:SchematicLGAD}
 \end{center}
\end{figure}

\IEEEPARstart{L}{ow} Gain Avalanche Diode (LGAD) is the choice for the High Granularity Timing Detector (HGTD)~\cite{HGTDtdr2020,Nicolo2018}, which is a thin N-on-P silicon sensor with a highly doped P+ layer inserted between N++ and p-type bulk, as shown in Fig.~\ref{fig:SchematicLGAD}. 
The P+ layer creates a high field and acts as a gain layer. 
The LGAD sensor has a good timing resolution and moderate position resolution ability. One of the important applications is the upgrade of the ATLAS experiments at the High-Luminosity Large Hadron Collider (HL-LHC)~\cite{HGTDtdr2020,Nicolo2018,HL-LHC}. The LGAD sensor is required to reach 70 ps timing resolution after irradiation up to \SI{2.5}{\times10^{15}~n_{eq}/\centi\metre^2}, and the collected charge is also required to be larger than 4 fC~\cite{HGTDtdr2020}. 

Ref.~\cite{Cartiglia2017,Kramberger2018,Zhao2018,Lange2017,Shi2020,Wiehe2021,Fan2020,limz2021} reported the radiation performance of HPK, CNM and IHEP-NDL LGAD sensors. The collected charge and timing resolution of the irradiated LGAD will deteriorate due to the acceptor removal mechanism~\cite{Moll2018,Ugobono2018, Kramberger2015,Ferrero2019}. FBK has designed gain layers with different carbon implanted and different depths to enhance the radiation hardness of LGADs~\cite{2017GPaternoster,2021Simone,Ferrero2019}. In this paper, the IHEP-IME designed sensors with the shallow carbon and deeper N++ layer to enhance the radiation hardness. The IHEP-IME sensors were irradiated by the neutron up to \SI{2.5}{\times10^{15}~n_{eq}/\centi\metre^2}. IHEP-IME LGAD sensors were designed by the Institute of High Energy Physics (IHEP) and fabricated by the Institute of Microelectronics (IME). 
%we tried to implant carbon in the P+ region to improve the irradiation hardness of LGAD sensors.
%The first version of the IHEP-IME LGAD has three wafers, wafer1 (W1), wafer7 (W7), and wafer8 (W8).
%Taking W7 as a reference, W1 has carbon implantation in the gain layer, and W8 uses a higher N+ layer implantation energy.
%This article will show the irradiation hardness performance of the three types of LGAD sensor after neutron irradiation up to \SI{2.5}{\times10^{15}~n_{eq}/\centi\metre^2}.

%%%%
%%%% 2.Parameters of IHEP-IME LGAD
%%%%
\section{Parameters of the IHEP-IMEv1 LGAD}
\label{sec_Properties}

\begin{figure}[!t]
\begin{center}
\includegraphics[scale=0.3]{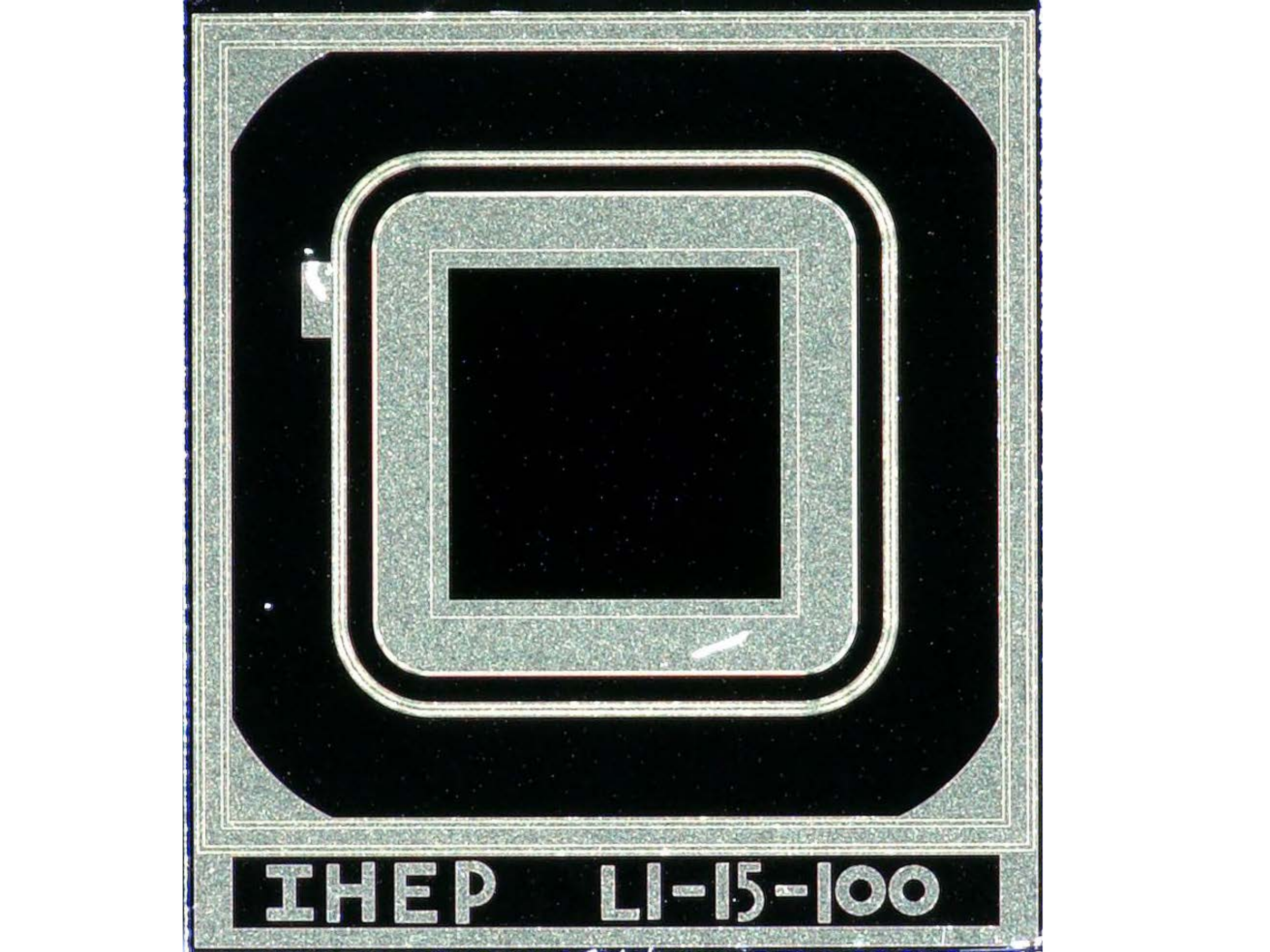} 
\caption{The layout of IHEP-IMEv1 LGAD sensor.}
\label{fig:Layout}
\end{center}
\end{figure}

The first batch of the IHEP-IME LGADs (IHEP-IMEv1) was fabricated on an 8-inch wafer. The wafer has a 50 \SI{}{\micro\metre} p-type epitaxial layer (active layer) and a 725 \SI{}{\micro\metre} P++ substrate. 
The IHEP-IMEv1 LGADs have three wafers, wafer1 (W1), wafer7 (W7), and wafer8 (W8). 
Table~\ref{tab:3wafers} shows the specific parameters of the three wafers.
Taking W7 as a reference, W1 has carbon implantation, and W8 has a deeper N+ layer because of higher implantation energy.
The resistivity of the active layer of the three wafers is about \SI{1000}{\Omega\cdot\centi\metre}. 
%IHEP-IMEv1 contains three types and is produced on three wafers respectively, of which wafer 7 (W7) is the basic design, W1 is designed with carbon doping based on W7, and W8 is based on W7 to increase the injection energy of the N+ layer.
The layout of the single pad IHEP-IMEv1 LGAD is shown in Fig.~\ref{fig:Layout}, the pad size is 1.3~mm~\si{\times}~1.3~mm.
The inner metal ring is a pad electrode, and the outer metal ring is a guard ring electrode.

\begin{table}
\centering
\caption{Implantation parameters of the W1, W7 and W8.}
\label{tab:3wafers}
\setlength{\tabcolsep}{5pt}    
	\begin{tabular}{ p{25pt} p{58pt} p{58pt} p{50pt}}
		\hline
		Sensors  & N++ layer energy (KeV) & P+ layer energy (KeV) & Carbon implantation   	\\
        \hline
		  W1   & 40  & 400 & Yes \\
         W7   & 40  & 400 & No \\
         W8   & 50  & 400 & No \\
		\hline
\end{tabular}
\end{table}

%%%%
%%%% 3.Neutron irradiation
%%%%
\section{Neutron irradiation}
\label{sec_irrad}

The IHEP-IMEv1 LGADs were irradiated with neutron at the Jozef Stefan Institute research reactor, which has been used successfully in the past decades to support sensor development~\cite{Snoj2012}.Three irradiation fluences are set, respectively \SI{0.8}{\times10^{15}}, \SI{1.5}{\times10^{15}} and \SI{2.5}{\times10^{15}~n_{eq}/\centi\metre^2}. After irradiation, the 50 \SI{}{\micro\metre} LGADs were annealed for 80 min at 60 \SI{}{\celsius}, which roughly simulates the room temperature annealing at the end of year shut-down period during the operation of the HL-LHC.

%%%%
%%%% 4.IV-CV
%%%%
\section{Effect of irradiation on capacitance and leakage current characteristics}
\label{IV-CV}

%%%%
\subsection{Current-Voltage(I-V)}

\begin{figure}[!t]
\begin{center}
\includegraphics[scale=0.4]{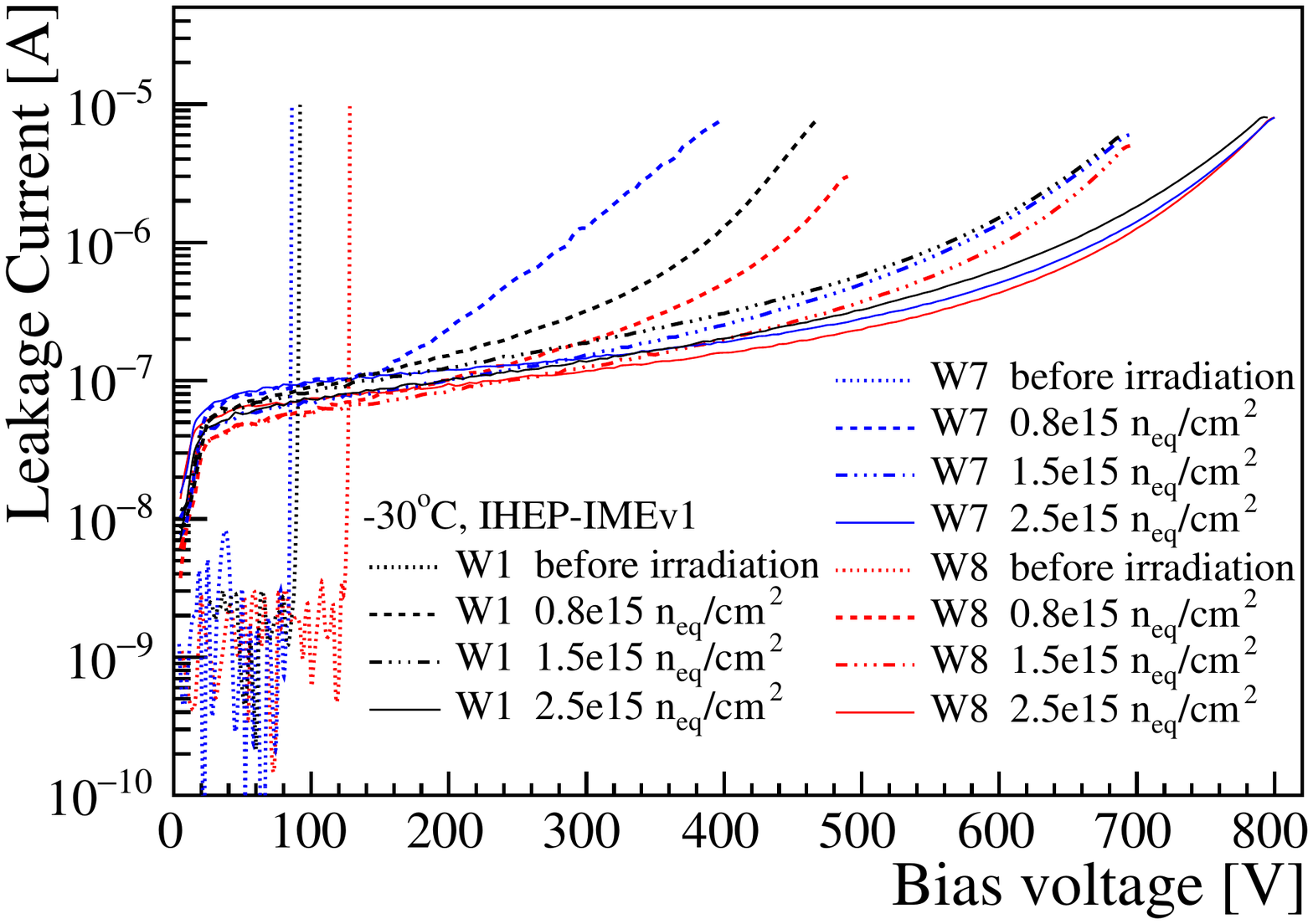} 
\caption{(color online) I-V curves of IHEP-IMEv1 LGAD sensors at -30\SI{}{\celsius}.}
\label{fig:IV}
\end{center}
\end{figure}

Figure~\ref{fig:IV} shows the leakage current of IHEP-IMEv1 LGAD sensors in a dark environment at -30\SI{}{\celsius}. 
Before irradiation, the leakage current of the three sensors at nA level, and breakdown at 86 V (W1), 92 V (W7), and 125 V (W8).
%While deeper N++ layer can lead to the breakdown voltage increasing. 
%With or without carbon implantation (W1 and W7), sensors show similar breakdown voltage.
The W1 with carbon implantation shows a little lower breakdown voltage.
The W8 with a deeper N+ layer has a higher breakdown voltage.
When the irradiation fluence up to \SI{2.5}{\times10^{15}~n_{eq}/\centi\metre^2}, the W1 has a slightly higher leakage current than W7 and W8, which is about 1~\SI{}{\micro\ampere} (59 \SI{}{\micro\ampere/\centi\metre^2}) at \SI{655}{V}. 
All the three sensors meets the HGTD requirements (\textless 125 \SI{}{\micro\ampere/\centi\metre^2}).

%%%%
\subsection{Capacitance-Voltage(C-V)}

\begin{figure}[!t]
\begin{center}
\includegraphics[scale=0.4]{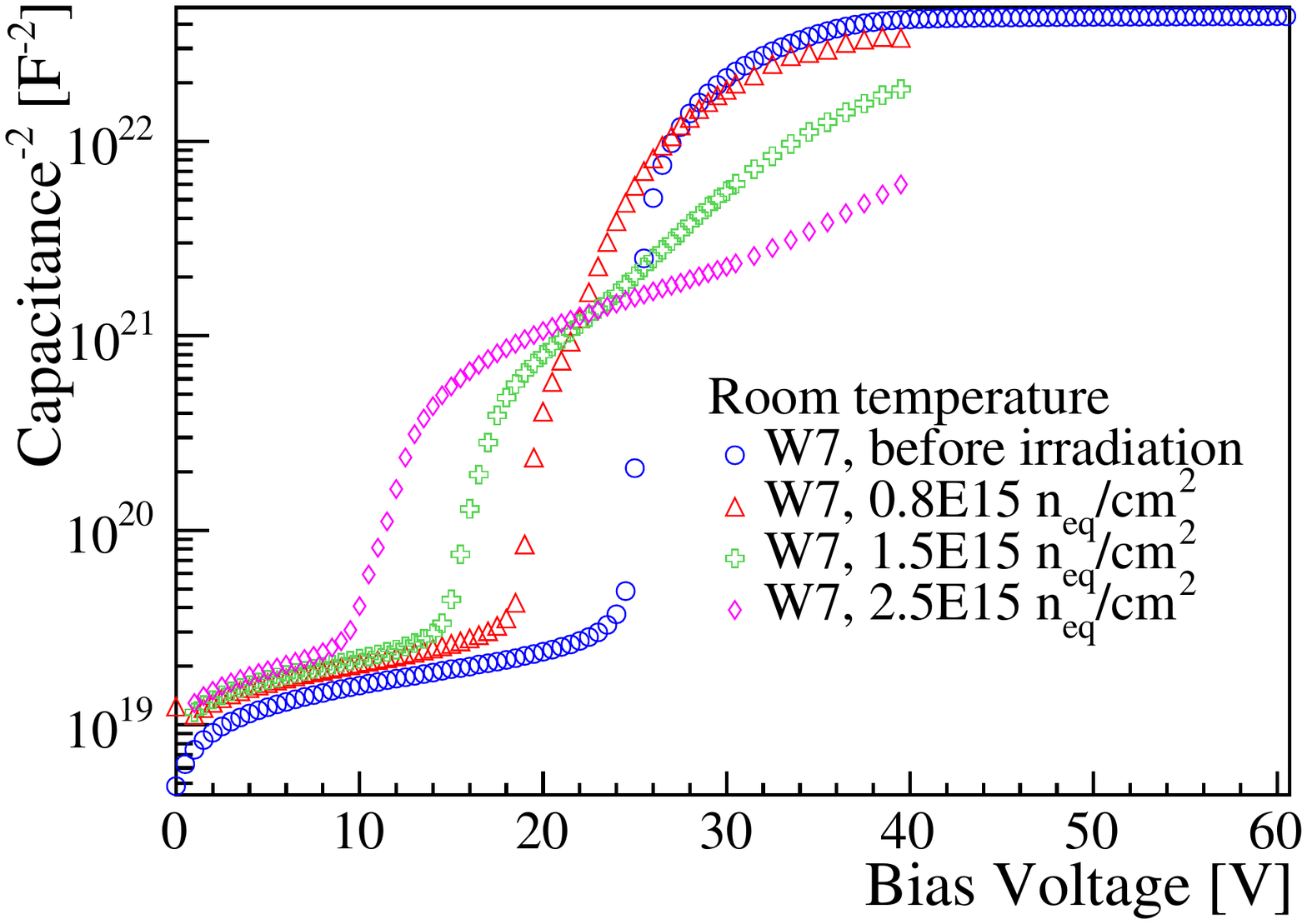} 
\caption{${1/C^{2}}$ as a function of bias voltage for IHEP-IMEv1 LGAD W7 sensors.}
\label{fig:C2_V}
\end{center}
\end{figure}

\begin{figure}[!t]
\begin{center}
\includegraphics[scale=0.4]{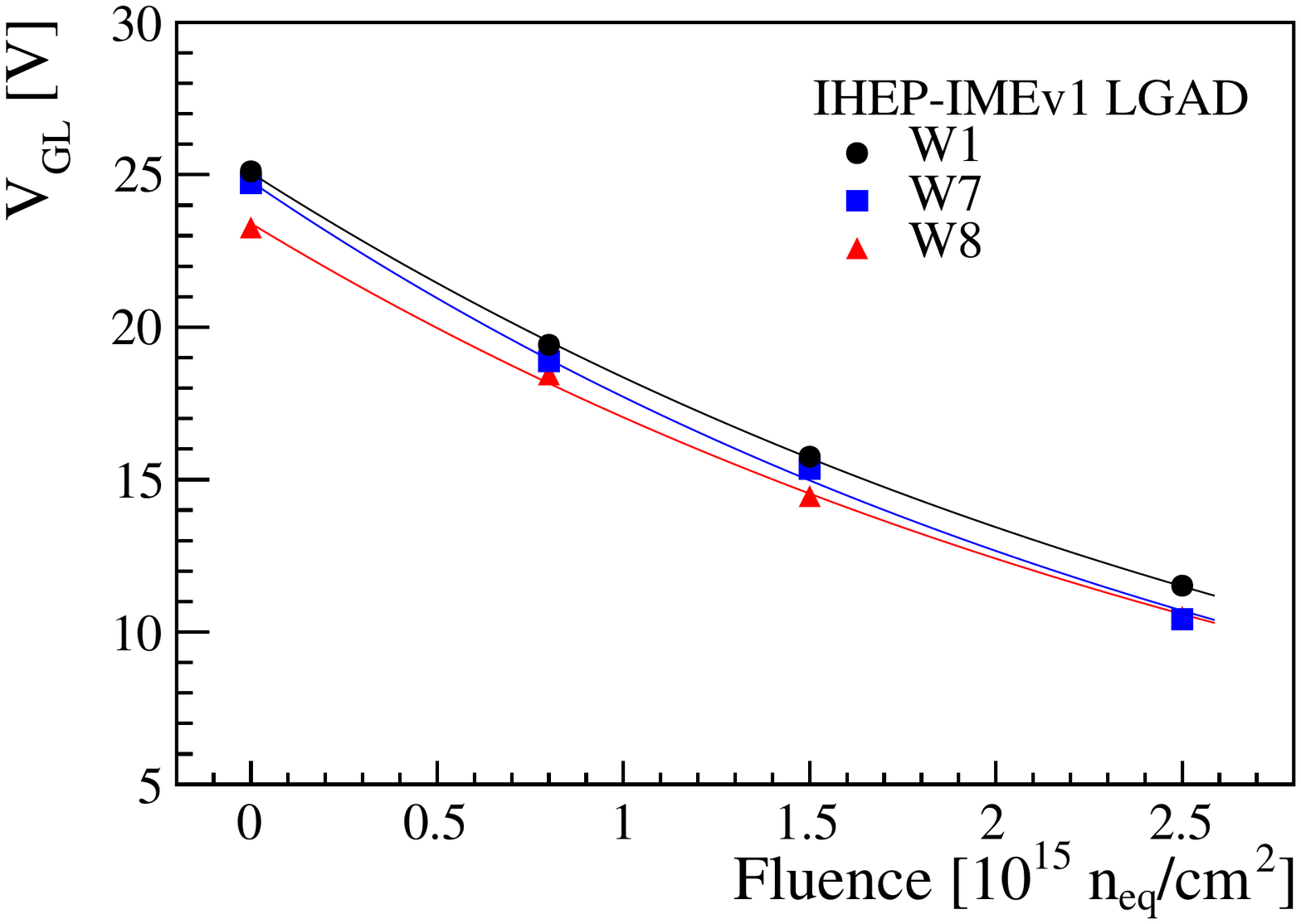} 
\caption{$V_{GL}$ as a function of irradiation fluence for IHEP-IMEv1 LGAD sensors.}
\label{fig:Vgl}
\end{center}
\end{figure}

LCR meter was used to test the C-V characteristics of the IHEP-IMEv1 LGAD sensors at room temperature. 
%The LCR meter frequency of the sensor after irradiation is set to 1 kHz, and the frequency of the sensor before irradiation is set to 10 kHz. 
Figure~\ref{fig:C2_V} shows the ${1/C^{2}}$ as a function of the bias voltage before and after irradiation sensors. 
As the irradiation fluence increased, the active acceptor of the gain layer is gradually removed, and the gain layer depletion voltage $V_{GL}$ is gradually reduced, as shown in Fig.~\ref{fig:Vgl}. The gain layer depletion voltage $V_{GL}$ is exponentially dependent on fluence ~\cite{ Kramberger2015}:
\begin{equation}
V_{GL} \approx V_{GL0} \cdot { exp}(-c \Phi_{eq}) 
\end{equation}
where $c$ is the removal constant, $V_{GL0}$ is the initial gain layer depletion voltage. The removal constant values of W1, W7 and W8  were found to be \SI{3.12}{\times10^{-16}}, \SI{3.36}{\times10^{-16}} and  \SI{3.17}{\times10^{-16}} respectively. W1 and W8 have smaller removal constants than W7.
%The W1 with carbon implantion shows better radiation hardness than W7. 

%%%%
%%%% 5.Low temperature beta telescope experiment
%%%%
\section{Effect of irradiation on the timing resolution and collected charges}
\label{beta}

%%%%
\subsection{Beta experiment setup}
\begin{figure}[htbp]
\begin{center}
\subfigure[]{\includegraphics[width=.31\textwidth]{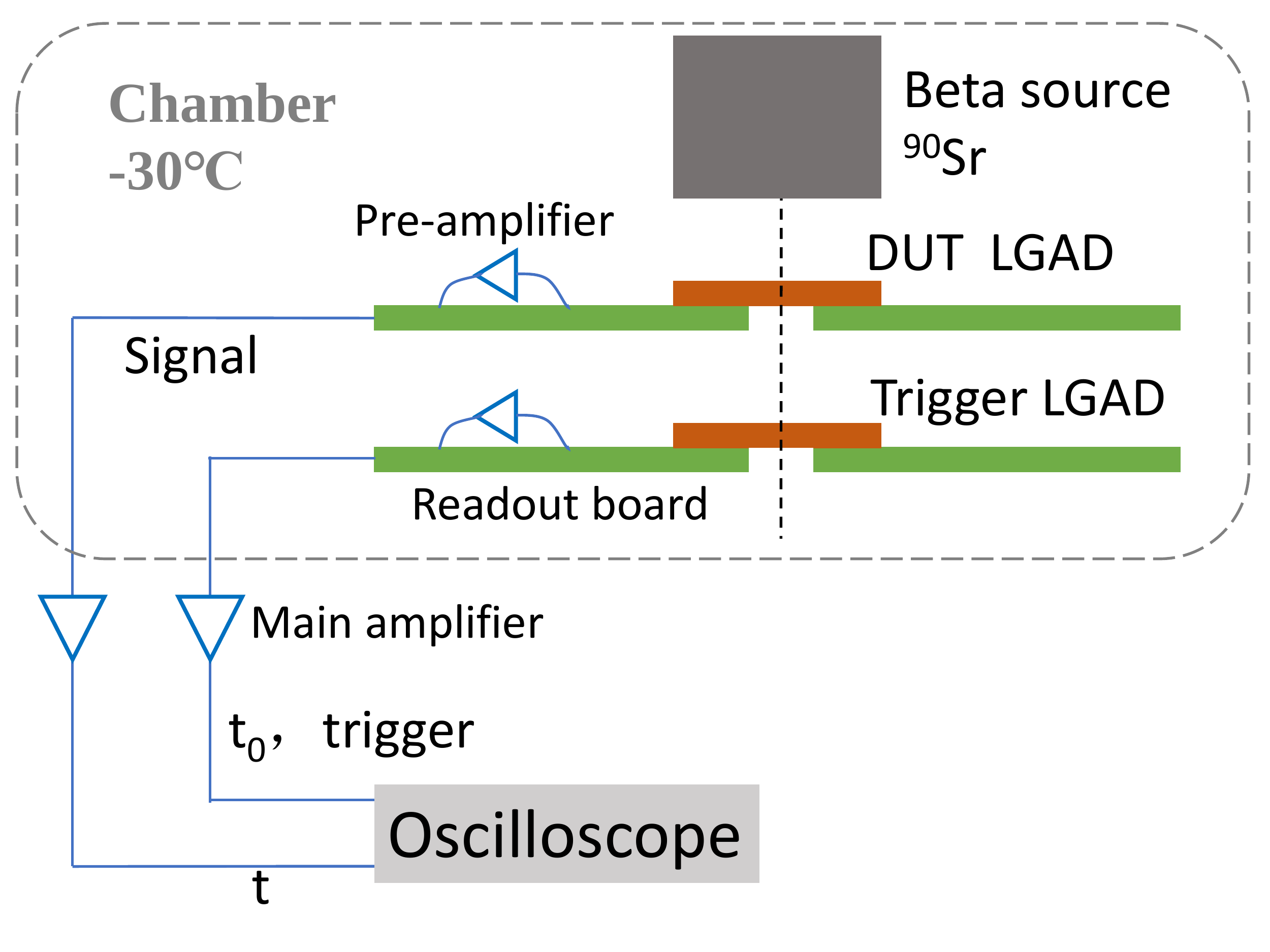} \label{fig:BetaSetup} }
\subfigure[]{\includegraphics[width=.3\textwidth]{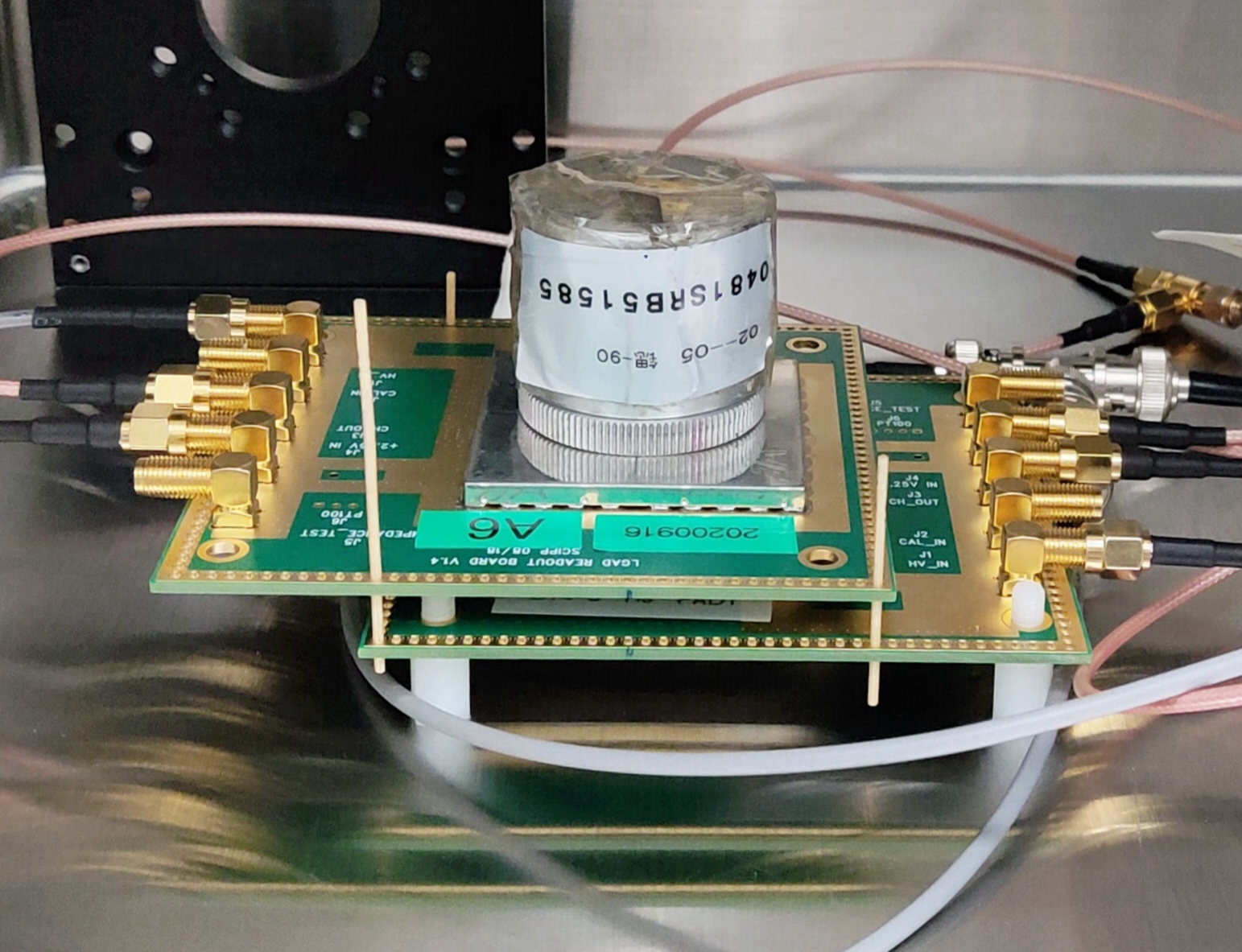} \label{fig:BetaSetup2} }
\caption{Beta telescope experiment setup for IHEP-IME sensors. (a) Schematic diagram. (b) Eexperimental picture.}
\label{fig:Beta}
\end{center}
\end{figure}

To study the timing resolution and collected charge, the LGAD sensors were tested with a beta source Sr-90 at -30\SI{}{\celsius}. The LGAD sensor is wire bonded to the readout board and the guard ring is grounded. The readout board is designed by the University of California Santa Cruz (UCSC)~\cite{Cartiglia2017}. It uses a broad band inverting trans-impedence amplifier of 470~\SI{}{\Omega}. This preamplifier is followed by an external second stage amplifier with a gain of 20~dB. Figure~\ref{fig:Beta} shows the beta telescope experiment setup, the lower sensor is used as the trigger for electrons signal in the beta telescope test. The signal pulses from both sensors are recorded by a digital oscilloscope with 40~GS/s sampling rate for offline analysis.

%%%%

%%%%
\subsection{Timing resolution}

%The timing resolution was obtained from the spread of the flight time $\Delta{t}$ of the electron passing through two LGADs.

%The CFD method can be used to calculate the flight time $\Delta{t}$. 
In the beta test, the timing resolution was obtained from the spread of the flight time $\Delta{t}$ of the electron passing through the Device Under Test (DUT) LGAD and the trigger LGAD.
The timing resolution of the DUT sensor could be calculated as the following equation:

\begin{equation}
\sigma_{DUT} = \sqrt{\sigma^2_{\Delta{t}} - \sigma^2_{trigger}} 
%\label{eq:thetaT}
\end{equation}
In our study, the IHEP-NDLv3 LGAD sensor is used as a trigger LGAD with a timing resolution of 31.2~ps at \SI{-30}{\celsius}~\cite{limz2021}. The detailed calculation for timing resolution was described in Ref.~\cite{Cartiglia2017}. 

\begin{figure}[!t]
\begin{center}
\includegraphics[scale=0.4]{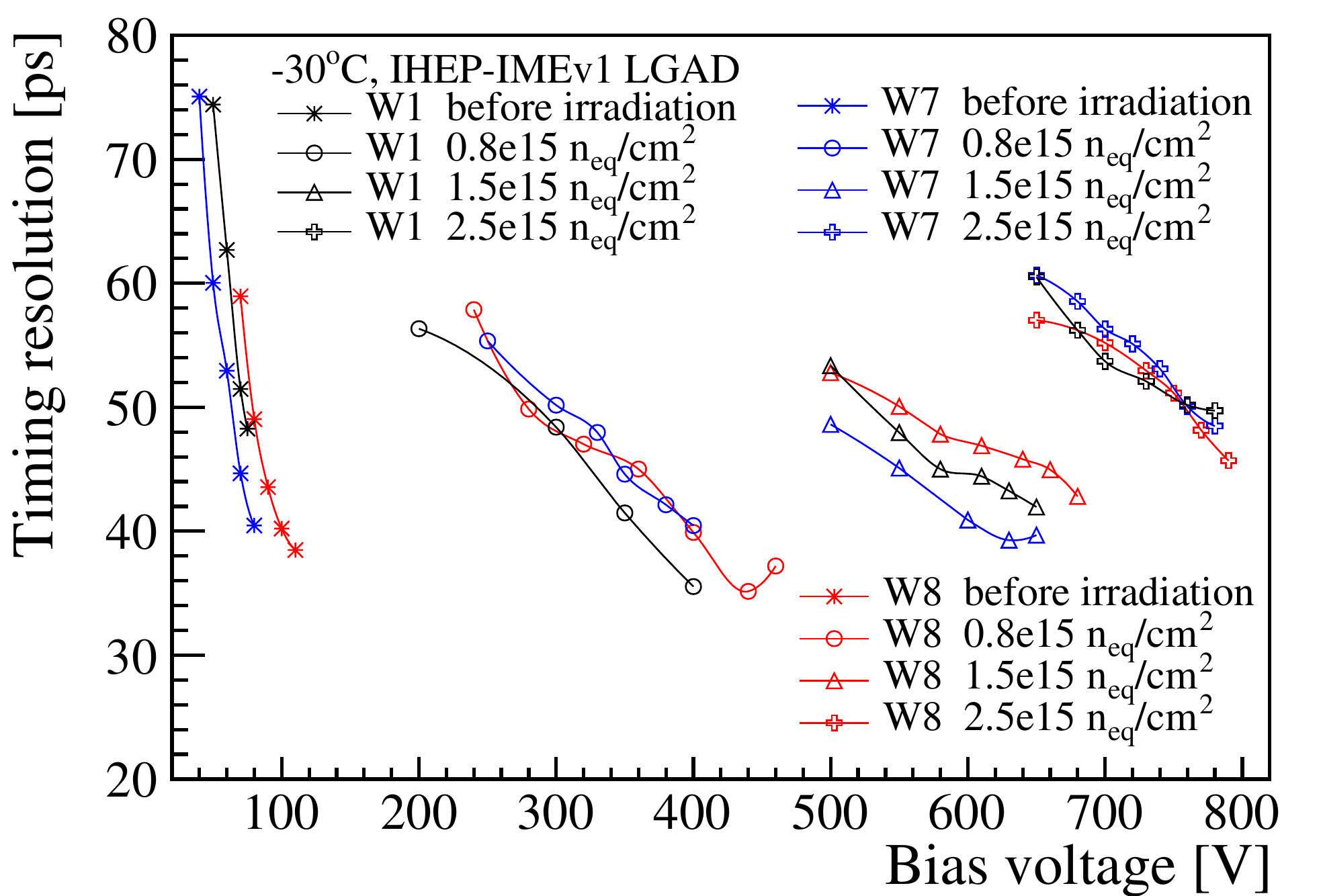} 
\caption{(color online) Time resolution as a function of bias voltage for IHEP-IMEv1 LGADs before and after irradiation at -30\SI{}{\celsius}}
\label{fig:Res}
\end{center}
\end{figure}

The timing resolution of the IHEP-IMEv1 LGAD sensors before and after irradiation is shown in Fig.~\ref{fig:Res}. 
Before irradiation, the best timing resolutions of the W8, W7, and W1 are 38 ps, 40 ps, and 48 ps. The timing resolution of the W1 with shallow carbon is worse than the others, which is probably due to the higher leakage current.

After irradiation, the best timing resolutions of the three sensors were obtained at the largest bias voltages which were the critical point before the large increase in dark count. At the irradiation points, \SI{0.8}{\times10^{15}~n_{eq}/\centi\metre^2}, the timing resolution of W1 is slightly better than those of the other two at most bias voltages. But at the irradiation point, \SI{1.5}{\times10^{15}~n_{eq}/\centi\metre^2}, the timing resolution of the W7 is the best. At \SI{2.5}{\times10^{15}~n_{eq}/\centi\metre^2}, the timing resolution of W1 is better than that of W7 at most bias voltage points but is worse than that of W8 and W7 after 760 \si{\volt}. To sum up, it's hard to say which sensor is better only in the point of the timing resolution. 
However, the performance of the three sensors is significantly different when focusing on the jitter contribution. 
%However, there's clearly different performances of the three sensors when we studied the jitter contribution. The detailed discussion could be seen in the following part.

%\SI{0.8}{\times10^{15}~n_{eq}/\centi\metre^2},\SI{1.5}{\times10^{15}~n_{eq}/\centi\metre^2} and \SI{2.5}{\times10^{15}~n_{eq}/\centi\metre^2}.

%jitter
The timing resolution mainly consisted of three items, the jitter item, the time walk item, and the landau item as the following equation,
\begin{equation} \label{eq:Timing}
\sigma^2_t = \sigma^2_{\mathrm{TimeWalk}} + \sigma^2_{\mathrm{Landau}} +  \sigma^2_{\mathrm{Jitter}} 
%\label{eq:sigmat}
\end{equation}

\begin{figure}[!t]
\begin{center}
\includegraphics[scale=0.4]{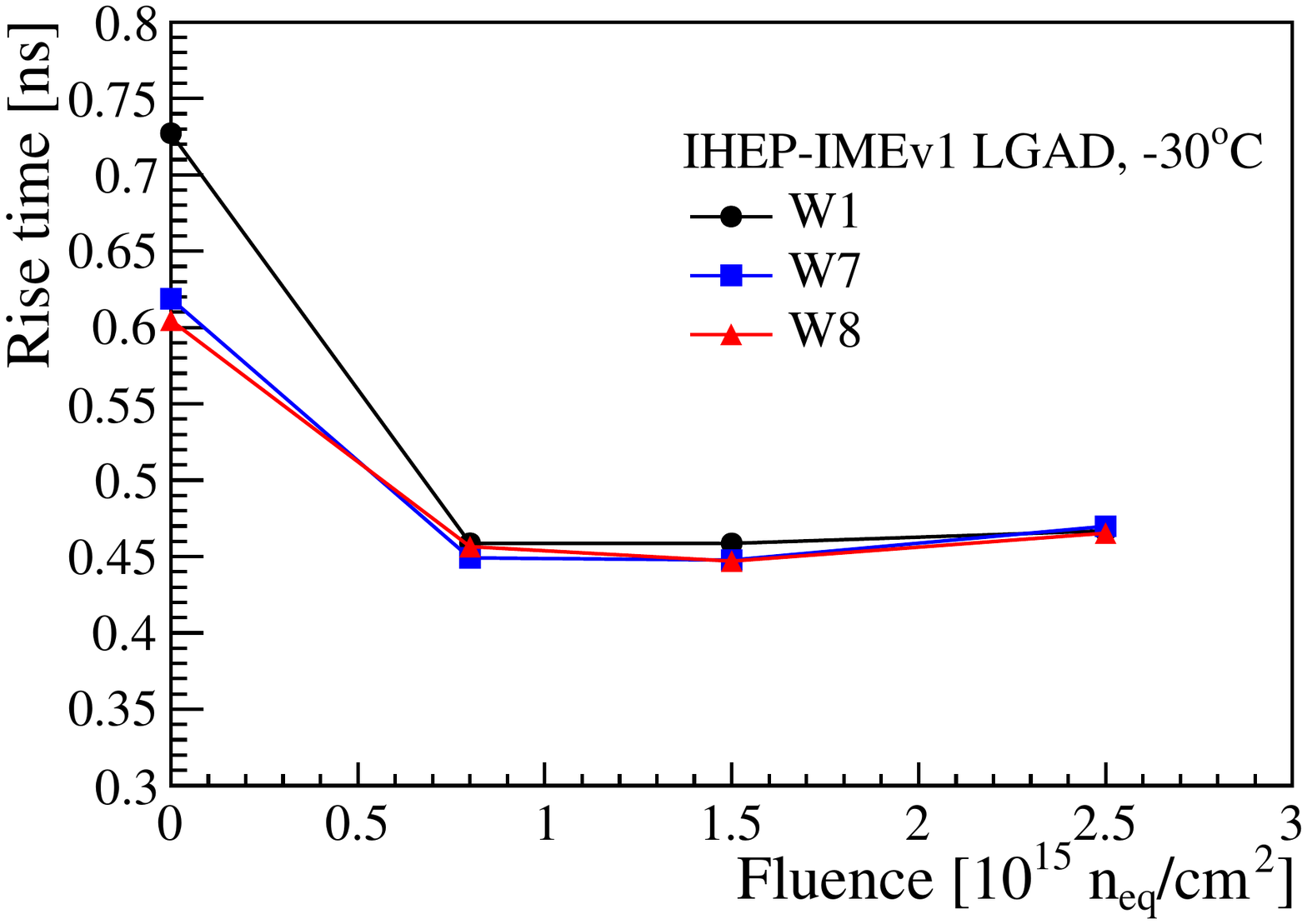} 
\caption{The rise time of the three sensors with with different irradiation fluences.}
\label{fig:riseTime}
\end{center}
\end{figure}

To correct for the time walk due to amplitude variations, the CFD algorithm can be used to calculate the time of electron hitting two LGADs and get the flight time $\Delta{t}$Ref.~\cite{Cartiglia2017}.
The jitter could be approximated by the following equation:
\begin{equation} \label{eq:Jitter}
    \sigma_{\mathrm{Jitter}} \approx \frac{t_{r}}{S/N}
\end{equation}
The $t_{r}$ is the rise time. The $S/N$ is the signal-to-noise ratio (SNR).
As shown in Fig.\ref{fig:riseTime}, the rise time didn't change too much with the increase of the irradiaition fluences after \SI{0.8}{\times10^{15}~n_{eq}/\centi\metre^2}.
Before irradiation, the rise time of different sensors are different when the sensors were biased to the same voltage. 
After irradiation, the rise time of the three sensors became almost the same even with different bias voltage which is 
 about 0.46 ns. The almost unchanged rise time is assumed to be due to the saturation of the electron drift velocity induced by the high bias voltage applied ($>$ 200 V) after irradiation. %The rise time didn't change with the different design of the three kinds of sensors. 
%The $S/N$ could also extract from the date. %As shown in Fig.\ref{fig:snr}, the SNR increased as the influence increased.

\begin{figure}[htbp]
\begin{center}
\includegraphics[scale=0.4]{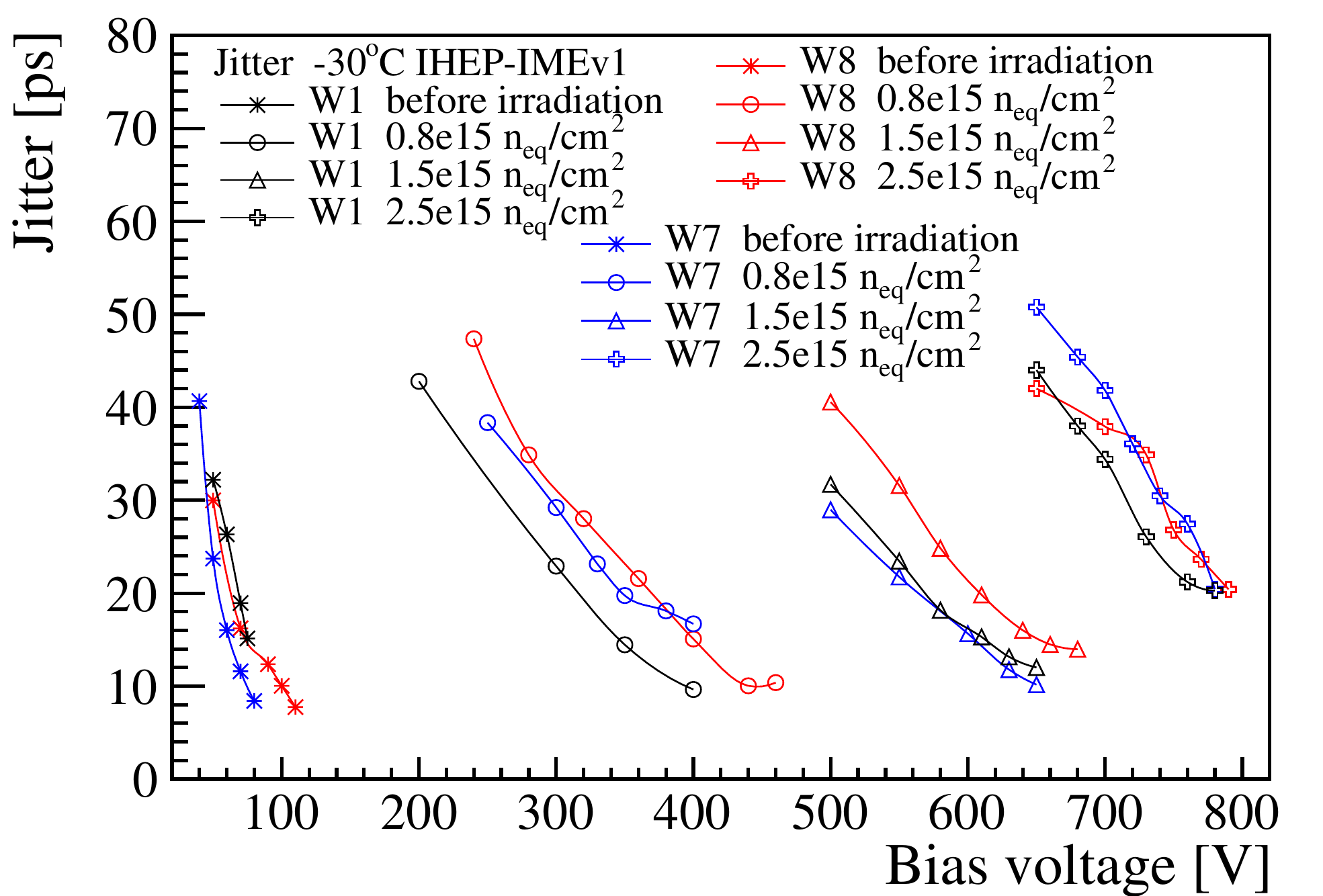} 
\caption{Jitter before and after irradiation }
\label{fig:jitter}
\end{center}
\end{figure}

According to the equation\ref{eq:Jitter}, the jitter could be calculated and the results are shown in Fig.\ref{fig:jitter}. After irradiation, the deterioration of the jitter of the W1 with shallow carbon is less than those of the W7 and W8. The deterioration of the jitter of W8 with a deeper N++ layer is the largest. 

% In the beta experiment, the signal shapes of the LGAD were recorded by a digital oscilloscope.
% Figure~\ref{fig:WW178} shows the normalized average waveforms of W1, W7, and W8 before and after irradiation.
% After irradiation, the overshoot of the sensors becomes severe than that before. The overshoot of W1 and w7 are better than that of W8.
%Figure~\ref{fig:riseTime} shows the rise time of the sensor of three wafers before and after irradiation.

%The highest working voltages of W1, W7, and W8 before irradiation are 75V, 80V, and 110V, respectively, and the drift velocity of electrons is not saturated at this time. W8 has a higher working voltage, and the electrons generated in W8 have a faster drift velocity, so it has a faster rise time. After irradiation, the working voltages of the three sensors were all higher than 200V, all reached the saturation drift velocity of electrons, reached the fastest rise time, and remained stable.

%\begin{figure}[htbp]
%\begin{center}
% \subfigure[]{\includegraphics[width=.3\textwidth]{Fig/NAWF_W1.pdf} \label{fig:WW1} }
% \subfigure[]{\includegraphics[width=.3\textwidth]{Fig/NAWF_W7.pdf} \label{fig:WW7} }
% \subfigure[]{\includegraphics[width=.3\textwidth]{Fig/NAWF_W8.pdf} \label{fig:WW8} }
% \caption{(color online) Average normalized pulse shapes for W1 (a), W7 (b) and W8 (c) with different irradiation fluences. }
% \label{fig:WW178}
% \end{center}
% \end{figure}

% \begin{figure}[htbp]
% \begin{center}
% \includegraphics[scale=0.4]{Fig/SNR_irradiation.pdf} 
% \caption{Signal-to-noise ratio (SNR) before and after irradiation }
% \label{fig:snr}
% \end{center}
% \end{figure}

%%%%
\subsection{Collected charge}

The collected charge for each signal pulse is calculated by dividing the integration of the pulse by the gain of amplifiers. The distribution of the charge collection is fited with the Landau-Gauss convolution to get the charge most probability value (MPV) for a certain bias. 

\begin{figure}[!t]
\begin{center}
\includegraphics[scale=0.4]{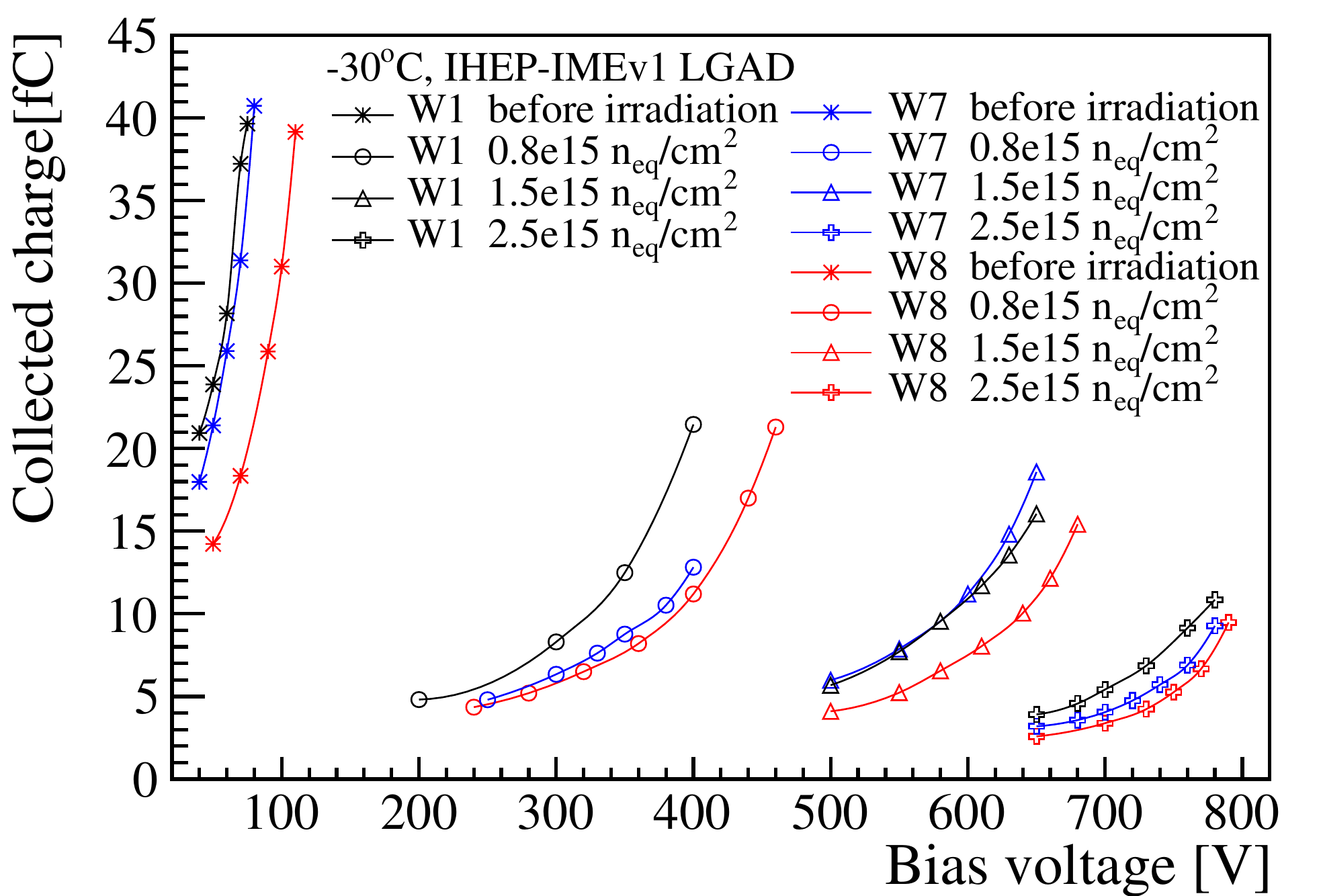} 
\caption{(color online) The collected charge as a function of bias voltage for IHEP-IMEv1 LGADs before and after irradiation at -30\SI{}{\celsius}}
\label{fig:Charge}
\end{center}
\end{figure}

The collected charge of the IHEP-IMEv1 sensors measured by the beta telescope before and after irradiation is shown in Fig.~\ref{fig:Charge}.  
Before irradiation, all three sensors have a very high collected charge, which is about 40 fC.
After irradiation, the collected charge of the three sensors could meet the HGTD requirements (\textgreater 4 fC after \SI{2.5}{\times10^{15}~n_{eq}/\centi\metre^2} irradiation fluence). The largest collected charges of all three sensors decreased. However, the deterioration of the collected charge of W1 is the samllest, while the deterioration of W8 is the largest. It indicates that the application of the shallow carbon could help increase the collected charge at the same irradiation fluence. While increase the injection energy of the N++ layer could decrease the collected charge after irradiation.

%After the irradiation fluence reaches \SI{2.5}{\times10^{15}~n_{eq}/\centi\metre^2}, the collected charge of W1 is higher than W7, and W8 is lower than W7.
%The W1 with carbon implantation has improved the irradiation hardness, while the W8 with a deeper N+ layer has not improved the irradiation hardness.
%The carbon implanted W1 in the gain layer has slightly better irradiation hardness.
%When the irradiation fluence up to \SI{2.5}{\times10^{15}~n_{eq}/\centi\metre^2}, 

%\begin{figure}[!t]
%\begin{center}
%\includegraphics[width=.6\textwidth]{Fig/Vat4fC.pdf} 
%\caption{Vat4fc}
%\label{fig:Vat4fC}
%\end{center}
%\end{figure}

\section{Operating bias voltage range after \SI{2.5}{\times10^{15}~n_{eq}/\centi\metre^2} irradiation}

After \SI{2.5}{\times10^{15}~n_{eq}/\centi\metre^2} irradiation, the sensors should operate on the bias voltage range which satisfied the HGTD project requirements (The timing resolution $<$ 70 ps; The collected charge $>$ 4 fC; The leakage current $<$ 125 \SI{}{\micro\ampere/\centi\metre^2} (2.1~\SI{}{\micro\ampere}) ).

Table~\ref{tab:Vat4fC} shows the operating bias voltage range of the three sensors.
The operating range of W1 with shallow carbon is 2 times wider than that of the W7 and 10 times wider than that of W8. Also, W1 could work at the lowest bias voltage compared with the other two sensors, which could help reduce the single event burnout observed in the testbeam.
At the point of the operating range, the shallow carbon could increase the radiation hardness while the deeper N++ will weaken the radiation hardness.

\begin{table}
\centering
\caption{The operating range of the W1,W7 and W8 after \SI{2.5}{\times10^{15}~n_{eq}/\centi\metre^2} neutron irradiation.}
\label{tab:Vat4fC}
\setlength{\tabcolsep}{5pt}    
	\begin{tabular}{p{25pt} p{60pt} p{60pt}}
		\hline
		Sensors   & Operating voltage range [\si{\volt}] & Operating range width [\si{\volt}]    	\\
        \hline
	      W1   & 655-705 & 50 \\
         W7   & 700-725 & 25 \\
         W8   & 725-730 & 5 \\
		\hline
\end{tabular}
\end{table}

%%%%
%%%% 6.Conclusion
%%%%
\section{Conclusion}
\label{sec_Conclusion}

The IHEP-IMEv1 LGAD sensors were designed by IHEP and fabricated by IME with 50 \SI{}{\micro\metre} epitaxial layer. 
We studied the radiation hardness of the three sensors with or without the shallow carbon and the deep N++ layer.
The three sensors were irradiated with different neutron fluence, \SI{0.8}{\times10^{15}~n_{eq}/\centi\metre^2}, \SI{1.5}{\times10^{15}~n_{eq}/\centi\metre^2} and \SI{2.5}{\times10^{15}~n_{eq}/\centi\metre^2}. The leakage current, timing resolution and collected charge are measured before and after irraidaiton.% The performance of the tree sensors meet the HGTD requirement(\textless 125 \SI{}{\micro\ampere/\centi\metre^2},\textgreater 4 fC and \textless 70~ps after \SI{2.5}{\times10^{15}~n_{eq}/\centi\metre^2} irradiation fluence).

The W1 with shallow carbon showed the most promising radiation hardness. After \SI{2.5}{\times10^{15}~n_{eq}/\centi\metre^2} irradiation, the collected charge is the largest at the same bias voltage and achieve 4 fC at the lowest bias voltage (655\si{\volt}). The operating voltage range is 655-705 \si{\volt} which is the widest and the lowest compared with the ranges of the other two designs. The timing resolution of W1 from the beta test is slightly lower than the other two designs at some bias voltages while the jitter contribution of W1 is the smallest at the same bias voltage. The acceptor removal rate is \SI{3.12}{\times10^{-16}} and the smallest. But the leakage current of W1 is the largest before and after irradiation due to the shallow carbon. The W8 with a deeper N++ layer showed the worst radiation hardness. After \SI{2.5}{\times10^{15}~n_{eq}/\centi\metre^2}irradiation, the collected charge is the smallest at the same bias voltage and achieve 4 fC at the highest bias voltage,725\si{\volt}. The operating bias voltage range is 725-730 \si{\volt} which is the narrowest and highest. The jitter contribution of W8 is the largest.

%Compared with W7, the W1 with carbon implantation has a higher collected charge and a wider optimal voltage range, the W8 with a deeper N+ layer has a lower collected charge and a narrow optimal voltage range. 
%Therefore, the carbon implantation significantly improves the radiation hardness of the IHEP-IMEv1 LGAD sensor.
%The W1 with carbon implantation has improved the irradiation hardness, while the W8 with a deeper N+ layer has not improved the irradiation hardness.
%In the next prototype run, IHEP will further optimize the carbon implantation design and process.

In summary, the leakage current, collected charge, and timing resolution of the three sensors of IHEP-IMEV1 all meet the HGTD requirements(125 \SI{}{\micro\ampere/\centi\metre^2}, \textgreater 4 fC and \textless 70 ps after \SI{2.5}{\times10^{15}~n_{eq}/\centi\metre^2} irradiation fluence). Before irradiation, the W1 sensor with shallow carbon showed the worst timing performance and moderate collected charge. However, after irradiation, the W1 sensor showed the best irradiation hardness and W8 with a deeper N++ layer showed the worst irradiation hardness. In our next prototype, we are going to vary the carbon dose and the depth to improve the irradiation hardness further.
%%%%
%%%% 
%%%%
\section*{Acknowledgement}

This work was supported by the National Natural Science Foundation of China (No.11961141014), the State Key Laboratory of Particle Detection and Electronics (No.SKLPDE-ZZ-202001), the Hundred Talent Program of the Chinese Academy of Sciences (Y6291150K2), the CAS Center for Excellence in Particle Physics (CCEPP), the Scientific Instrument Developing Project of the Chinese Academy of Sciences (No.ZDKYYQ20200007). 
%Thanks to the Institute of Microelectronics for the detector production. 

%%%---
\appendices

\section*{References}

\def\refname{\vadjust{\vspace*{-1em}}} %Please don't do this in a real paper.

%\subsection*{Basic format for books:}


%\begin{thebibliography}{00}


%\bibitem{}

%\end{thebibliography}

%\section*{Acknowledgment}

\end{document}